\documentclass[useAMS,usenatbib]{mn2e}
\usepackage{epsf,palatino,url,graphicx,wrapfig,array,setspace,amsmath,amssymb,fancyhdr,multirow,lscape,appendix,rotating}
\def\apj{ApJ }
\def\aap{A\&A }

\def\pasp{PASP }

\def\mnras{MNRAS }

\def\apjs{ApJS }
\def\aj{AJ }
\def\apjl{ApJL }
\def\aj{AJ }

\def\apjs{ApJS }

\title[The Search for High-Mass X-ray Binaries in the Phoenix Dwarf Galaxy]{The Search for High-Mass X-ray Binaries in the Phoenix Dwarf Galaxy}
\author[E. S. Bartlett et al.]{E. S. Bartlett$^{1}$\thanks{E-mail:
e.s.bartlett@soton.ac.uk (ESB)}, M. J. Coe$^{1}$, F. Haberl$^{2}$, V. A. McBride$^{1}$ and R. H. D. Corbet$^{3}$ \\
$^{1}$School of Physics and Astronomy, University of Southampton, Highfield, Southampton, SO17 1BJ, United Kingdom\\
$^{2}$Max-Planck-Institut f\"ur extraterrestrische Physik, Giessenbachstra\ss{}e, 85748, Germany\\
$^{3}$University of Maryland Baltimore County, X-ray Astrophysics Laboratory, Mail Code 662, NASA Goddard Space Flight Center, Greenbelt, MD 20771, USA}
\begin{document}

\date{Accepted 2012 February 20.  Received 2012 February 4; in original form 2011 March 4}

\pagerange{\pageref{firstpage}--\pageref{lastpage}} \pubyear{2012}

\maketitle

\label{firstpage}

\begin{abstract}
We report on the first X-ray images of the Phoenix dwarf galaxy, taken with \emph{XMM-Newton} in July 2009. This local group dwarf galaxy shares similarities with the Small Magellanic Cloud (SMC) including a burst of star formation $\sim$50~Myr ago. The SMC has an abundance of High Mass X-ray Binaries (HMXBs) and so we have investigated the possibility of an HMXB population in Phoenix with the intention of furthering the understanding of the HMXB-star formation rate relation. The data from the combined European Photon Imaging Cameras (EPIC) were used to distinguish between different source classes (foreground stars, background galaxies, AGN and supernova remnants) using EPIC hardness ratios and correlations with optical and radio catalogues. Of the 81 X-ray sources in the field of view, six are foreground stars, four are galaxies and one is an AGN. The remaining sources with optical counterparts have log($\frac{f_X}{f_{opt}}$) consistent with AGN in the local universe. Further investigation of five sources in the field of view suggests they are all background AGN. Their position behind the gas cloud associated with Phoenix makes them a possible tool for further probing the metallicity of this region. We find no evidence for any HMXBs in Phoenix at this time. This rules out the existence of the X-ray persistent supergiant X-ray binary systems. However the transient nature of the Be/X-ray binaries means we cannot rule out a population of these sources but can conclude that it is not extensive.
\end{abstract}

\begin{keywords}
galaxies: individual: Phoenix dwarf - X-rays: galaxies - X-rays: binaries
\end{keywords}

\section{Introduction}

The Phoenix dwarf galaxy was discovered in 1976 by \citeauthor{Schust76} who described it as a very distant ($\sim$100~kpc) globular cluster. It was later identified as a dwarf irregular galaxy (dIrr) by \citet{Cant77}. These galaxies are H\textsc{i} rich and show obvious signs of recent star formation. Despite this classification, Phoenix has characteristics of a dwarf spheroid galaxy (dSph) and it has been suggested that it belongs to an intermediate class of galaxy between the dwarf spheroidal and dwarf irregular, along with three other dwarf galaxies (Pegasus, Pisces (LGS3) and Antila). The distance to Phoenix has been determined to be $\sim$420~kpc using a variety of different methods and data sets (for example see \citealt{Rydt91,Mart99} and, more recently, \citealt{Menz08}) placing Phoenix well within the Local Group.

The H\textsc{i} gas surrounding Phoenix was mapped by Young et al. \citeyearpar{Young97,Young07} using the VLA. They identify several regions in the immediate vicinity with a variety of shapes and velocities. One such cloud, located $\sim$5\arcmin south west of the main stellar body, has been unequivocally associated with Phoenix based on the excellent agreement of the radial velocities of the stars, obtained from stellar spectra \citep{Irwin02} from the VLT, and the velocity of the H\textsc{i} cloud. \citet{Young07} recognise that this offset suggests that Phoenix could offer a unique opportunity to study the possible mechanisms responsible for gas removal in dwarf galaxies, transforming a gas-rich dIrr into a gas-poor dSph.

Extensive optical data of the central region of Phoenix also exists: the Wide Field Planetary Camera (WFPC2) onboard \emph{HST} has imaged the central field of the galaxy with both F814W and F555W filters (for example see \citealt{Holtz00} and \citealt{Hid09}, hereafter H09). \citet{Young07} use both of these data sets to derive separate star formation histories for the eastern and western sides of Phoenix. They report an asymmetry in the star formation rate across the face of the galaxy, in agreement with \citet{Mart99}. The western side of Phoenix displays evidence for strong episodes of star formation at 180 and 40~Myr ago whilst the eastern side shows evidence for more continual star formation from 250 to 50~Myr ago. They note that this is consistent with the H\textsc{i} location, offset to the west of the galaxy's optical emission. These epochs of star formation are similar to those for another dIrr galaxy: the Small Magellanic Cloud (SMC). \citet{Zaritsky03} report that the SMC had ``bursts'' of star formation at 2.5, 0.4 and 0.06~Gyr in the past. In fact, the Phoenix dwarf galaxy shares many other common features with the SMC, including its low metallicity (Z=0.004 for the SMC, \citealt{Russell92}; Z=0.0015 for Phoenix, H09).

The SMC is $\sim$100 times as massive as Phoenix and host to an unusually high number of High-Mass X-ray Binaries (HMXBs) \citep{Coe05}. These are stellar systems in which a compact object, usually a neutron star, accretes material from a more massive donor - either a supergiant or early-type star, both of OB class. Supergiant systems are a persistent source of X-rays with luminosities, $L_X$, $10^{36}$--$10^{38}$ ergs~s$^{-1}$. Be/X-ray binary systems have two types of outburst associated with their X-ray emission. Type I outbursts have L$_X$ in the range $10^{36}$-$10^{37}$ ergs~s$^{-1}$ and occur periodically, around the time of the periastron passage of the neutron star, Type II outbursts reach higher luminosities, L$_X \geq10^{37}$ ergs~s$^{-1}$ and show no correlation with orbital phase \citep{Stella86}. A comparison with the Milky Way suggests we see a factor of $\sim$50 more HMXBs in the SMC than we would expect to observe, based on the mass ratio of the two galaxies. 

The stellar winds of massive stars are known to be line driven \citep{Lucy70} and so a lower metallicity will lead to less mass being lost from the compact object progenitor. This in turn is expected to affect the number of systems that survive the initial supernova explosionand go on to form HMXBs. \citet{Dray06} use simulations to demonstrate that this is indeed the case but are unable to replicate the the number and period distribution of the SMC HMXBs with a low metallicity environment alone. It has been known for some time that HMXBs are strong tracers of recent star formation episodes in a galaxy. It is currently thought that an increase in star formation, possibly caused by tidal interactions with the Milky Way and/or the Large Magellanic Cloud (LMC) \citep{Zaritsky03} along with the low metallicity of the SMC have given rise to the large number of HMXBs in the SMC. \citeauthor*{Grimm03} (2003, hereafter G03) have quantified this relationship for sources with X-ray luminosities in excess of $10^{38}$~ergs~s$^{-1}$. However Be/X-ray binaries, which are the most numerous subclass of HMXB \citep{Liu06}, do not reach these luminosities. One of the reasons for choosing such a high threshold luminosity was to avoid contamination from any low-mass X-ray binaries (LMXBs) present in any of the galaxies studied. \citet{Gilf04} has shown that the number of low-mass X-ray binaries (LMXBs) scales linearly with the mass of the host galaxy, therefore the low mass Local Group dwarf galaxies like Phoenix ($3.3\times10^{7}$~M$_\odot$, \citealt{Mateo98}) which are too low mass to expect LMXBs are ideal for probing the HMXB--star formation relation in the low luminosity limit. By studying any population present we hope to further understand factors that can influence star formation in such diverse environments.

Predicting the number of HMXBs cannot be done with great accuracy as the effects of all the influencing parameters are not fully understood. If we assume that the LogN-LogS relationship reported by G03 for HMXBs remains valid down to X-ray luminosities of $5\times10^{35}$~ergs~s$^{-1}$, the typical luminosity of the lower type I outbursts seen in the SMC \citep{Haberl08b}, equation (7) of G03 and the average star formation rate (SFR) of Phoenix ($1\times10^{-3}$~M$_\odot$~yr$^{-1}$, calculated from data in H09) predicts 0.14$\pm$0.02. However, this apparently precise number could be significantly affected by many factors including the metallicity values. The same equation predicts $\sim$20 X-ray sources with $L_X > 5\times10^{35}$~ergs~s$^{-1}$ (adopting the same value for the SFR as G03, 0.15~M$_\odot$~yr$^{-1}$, taken from \citet*{Yoko00}) for the SMC. We currently know of $\sim$60 HMXBs in the SMC\citep{Liu05}, of which $\sim$50 are X-ray pulsars, and more are being discovered every year (e.g. SXP1062, \citealt*{Henault11}). Using a SFR of 0.2~M$_\odot$~yr$^{-1}$ for the LMC \citep{Zaritsky09} leads to a predictions of $\sim$30 sources in the LMC with this luminosity constraint. There are currently 36 known HMXBs in the LMC, including several supergiant systems and black hole candidates.

The reason behind the apparent discrepancy in the SMC could lie with both the activity cycles of these HMXBs and the details of their star formation histories. The number of HMXBs in a galaxy will be more strongly linked to the recent star formation episodes, rather than the average star formation of the galaxy over its lifetime. The SFR of the more recent episodes of star formation (40-60~Myr ago) in Phoenix and the SMC are very similar. The SMC varies from $8.1\pm1.4\times10^{-5}$ to $1.5\pm0.6\times10^{-5}$~M$_\odot$~yr$^{-1}$arcmin$^{-2}$ \citep{Ant10} depending on the area of the SMC, whilst the most recent epoch of star formation in Phoenix had a SFR of $2.5\pm1.8\times10^{-5}$~M$_\odot$~yr$^{-1}$arcmin$^{-2}$ \citep{Young07}. It is only by probing galaxies across a broad spectrum of mass, star formation history, and chemical composition that we can establish both trends in the HMXB populations of these galaxies and understand the significance of these objects as tracers of star formation.

\section{Observations and Data Reduction}

The first X-ray images of the Phoenix dwarf galaxy were taken with \emph{XMM-Newton} during the satellite revolution \#1754 on July 7, 2009. This paper will discuss the data
\begin{table}
\centering
\label{observations}
\begin{tabular}{cccccc}
\hline\hline
Camera & Filter & Read out & \multicolumn{2}{c}{Observation} & Exp. \\
& & Mode & Start & End(UT) & (ks) \\
\hline
MOS1/2 & Medium & FF, 2.6~s & 08:13 & 21:11 & 46.6 \\
pn & Thin1 & FF, 73~ms & 08:36 & 21:06 & 45.0 \\
\hline
\end{tabular}
\caption{\emph{XMM-Newton} EPIC observations of the Phoenix galaxy on July 7, 2009}\label{table:observations}
\end{table}
collected with the EPIC MOS \citep{Turner01} and pn \citep{Struder01} detectors. All detectors were operating in Full Window modes, giving a field of view of $\sim$ 30' diameter. Table \ref{table:observations} summarises the details of the EPIC observations. The data were processed using the \emph{XMM-Newton} Science Analysis System v9.0 along with software packages from FTOOLS v6.8.

The MOS and pn observational data files were processed with \textsf{emproc} and \textsf{epproc} respectively. Periods of high background activity were screened by removing any times when the count rate above 10~keV was $>$0.8 cts s$^{-1}$ for the MOS detectors and \textgreater2.0 cts s$^{-1}$ for the pn detector. The filtered event files were then split into five energy bands: (0.2--0.5), (0.5--1.0), (1.0--2.0), (2.0--4.5) and (4.5--12.0) keV to identify and discriminate between hard and soft sources in the field of view. For the pn detector, we selected only the ``single'' (PATTERN=0) pixel event patterns in the range 0.2--0.5~keV, for all the other bands ``single and double'' (PATTERN$\leq$4) pixel events were accepted. For the MOS, ``single'' to ``quadruple ''(PATTERN$\leq$12) pixel events were selected. 

For all detectors we created images, background and exposure maps in each of the five energy bands. A box sliding detection was performed simultaneously on all 15 images (5 energy bands for each of the MOS1, MOS2 and pn cameras) with the task \textsf{eboxdetect} as a precursor to the maximum likelihood fitting task \textsf{emldetect}. The maximum likelihood fitting yielded 90 sources in the field of view, each with a total detection likelihood larger than 10. This corresponds to a probability of a source being a spurious detection of $\sim5\times10^{-5}$ in a single image, and $3\times10^{-6}$ for the case of simultaneously using 15 images\footnote{http://xmm.esa.int/sas/current/doc/emldetect/index.html}. The data were screened by eye to remove obvious false detections caused by instrumental effects, resulting in a final list of 81 sources.

To convert source count rates into flux values, the energy conversion factors (ECFs) from the \emph{XMM-Newton} Serendipitous Source catalogue \citep{2XMM} were used. Whilst these ECFs do assume a spectrum typical for an AGN with low absorption, \citeauthor{Pietsch04} (2004, hereafter P04) show that the ECFs calculated for a typical supernova remnant (an absorbed 1~keV thin thermal spectrum) and supersoft source (absorbed 30~eV black body spectrum) in M33 only vary from those calculated for a typical hard source by about 20\% for both EPIC detectors. The flux values calculated using the ECFs are used in source classification (see Section 3 of this work) over a broad energy range (0.2--4.5~keV, bands 1--4) to be consistent with the findings of \citet{Macca88} and the method of P04. A large deviation from a power law spectrum would be required for these results to be inadequate over such a range. The flux values for band 5 are calculated, but are not used in the source classification.

Four of the X-ray brightest sources (for exact details on which sources, see Section 5) were identified with \textgreater500 counts in the full energy range (0.2--12.0~keV) across all three detectors and were subject to further spectral analysis. EPIC spectra were extracted for the pn (PATTERN$\leq$4) and both MOS (PATTERN$\leq$12) detectors, bad pixels and columns were disregarded (FLAG=0) and intervals of high background were removed. All four sources were close to the centre of the field of view and as such were extracted with circular source regions with radius 20\arcsec.

For the MOS detectors, the background spectra were taken from an annulus surrounding the source extraction region with an outer radius of 60\arcsec. Where this was not possible, due to the proximity of other sources in the field, regions on the same chip which contained no sources were identified. Light curves and spectra were extracted (with radius 60\arcsec) from these regions and examined to confirm that they were statistically identical. The spectrum from the closest region to a source was used as the background spectrum. For the pn detector, regions on neighbouring chips which contained no sources were identified and again were confirmed to be statistically identical. As with the MOS detectors, the spectrum from the closest region to a source was used as the background spectrum. The area of source and background regions were calculated using the \textsf{backscal} task. Response matrix files were created for each source using the task \textsf{rmfgen} and \textsf{arfgen}. 

\section{Source Classification}

\begin{figure}
 \begin{center}
  \includegraphics[height=90mm,angle=90]{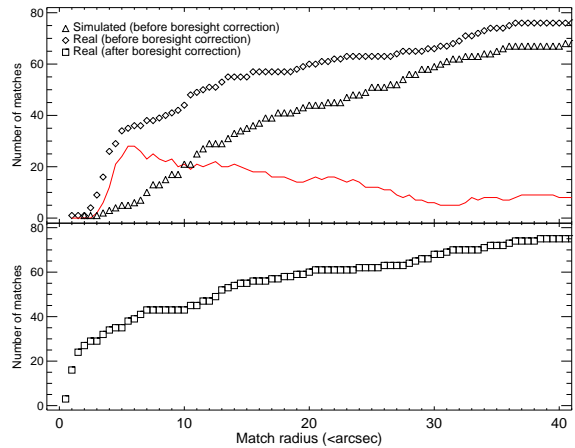}
 \end{center}
\caption{The top panel shows the cumulative number of matches as a function of search radius between the LGGS catalogue and the positions of the X-ray sources (diamonds). The number of matches between the X-ray positions and the simulated optical catalogue are also plotted (triangles). The solid red line is the difference between these two values. The bottom panel shows the number of matches between the LGGS catalogue and the boresight corrected X-ray positions are shown for comparison (squares).}\label{fig:cumulative}
\end{figure}

The positions of the 81 X-ray sources in Phoenix were cross-correlated with the Local Group Galaxy Survey (LGGS, \citealt{LGGS}) catalogue and the separation to the nearest optical match calculated. This was compared with the number of matches found with a simulated catalogue. The simulated positions were generated by independently reassigning both the right ascensions and declinations of all the optical sources. Figure \ref{fig:cumulative} shows the results of the correlations as a function of search radius. The solid line shows the difference between the number of X-ray sources matched with the real catalogue and the number matched with the simulated catalogue. The radius at which the number of real matches increases at the same rate as the number of simulated matches, i.e. the radius at which the solid line plateaus, can be considered as the maximum search radius. At radii greater than this value, all matches with the real catalogue are false associations. Our figure indicates that this occurs at about 6\arcsec.

\begin{figure*}
\centering
 \includegraphics[height=225mm]{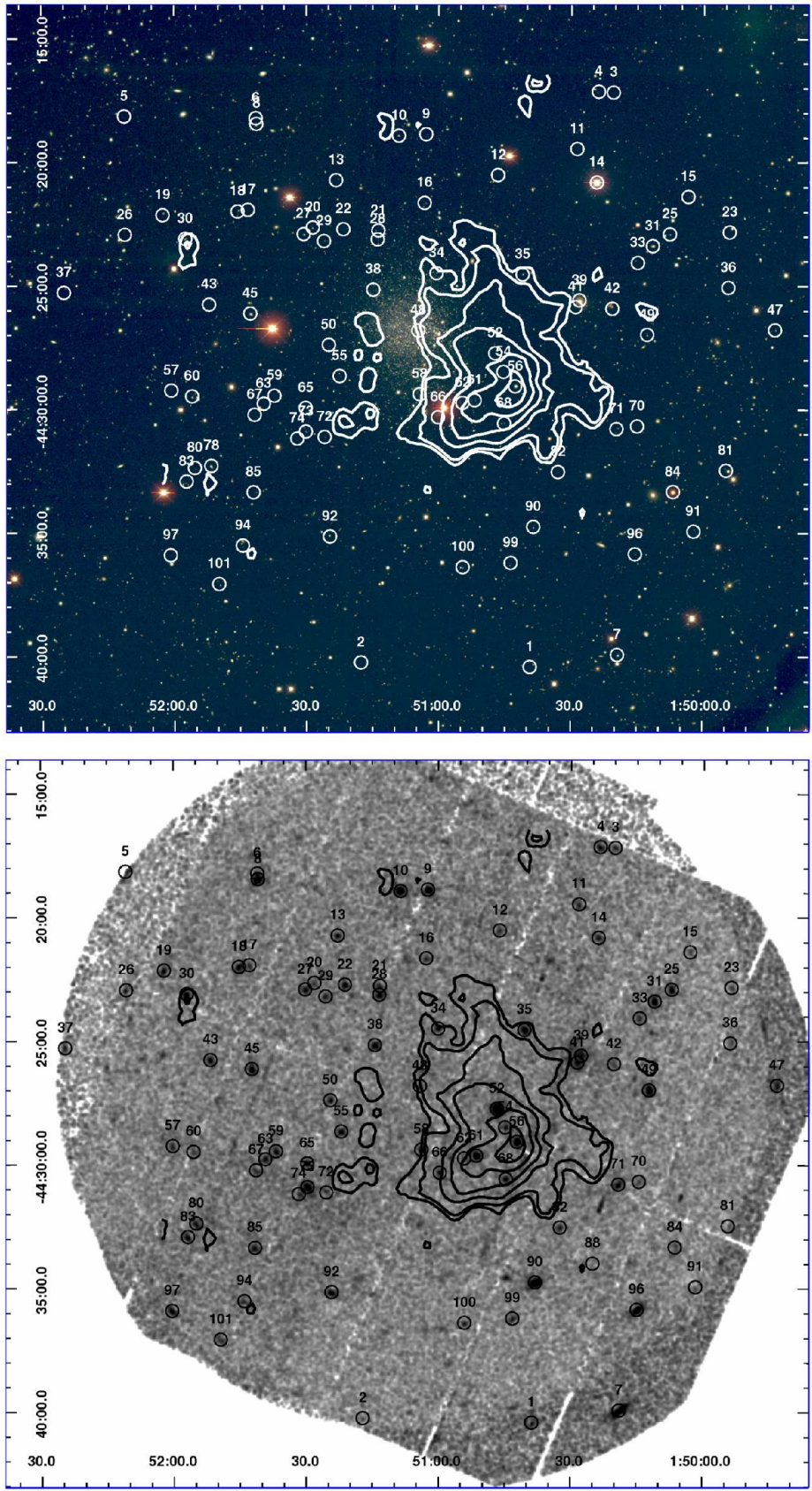}
\caption{Top panel shows the optical Local Group Galaxy Survey (LGGS) image, the bottom panel shows the combined 0.2-4.5~keV EPIC data. H\textsc{i} contours from \citet{Young07} are overlaid on both images. The contour levels correspond to column densities $(0.5,1,2,3,4,5)\times10^{19}$ cm$^{-2}$. The circles are the positions of the sources detected by the box sliding and maximum likelihood detection analysis, they are not indicative of the size of the error circles and are illustrative only. }\label{fig:Phoenix}
\end{figure*}

The source classification described in this section was performed in two stages. The SIMBAD and NED archives and the LGGS and United States Naval Observatory-B1.0 (USNO B1.0, \citealt{Monet03}) catalogues were searched for correlations around the X-ray source positions. The preliminary classification used a search radius of up to 6\arcsec for all sources. This was primarily to identify galaxies and AGN for an astrometric boresight correction. Foreground stars were purposely not used when performing the boresight correction due to the possibility of proper motion. The four galaxies and possible AGN listed in section 4.2 identified from the SIMBAD and NED databases were used along with the SAS task \textsf{eposcorr} to obtain accurate positions for all our sources. The final boresight correction was 4.7$\pm$1.5\arcsec. The bottom panel of Figure \ref{fig:cumulative} shows the number of matches between the corrected X-ray positions and the LGGS catalogue for comparison. The gradient of the curve decreases dramatically after 2\arcsec indicating that our boresight correction has been successful and that the error on our X-ray source positions are dominated by the error the required co-ordinate shift. This error is added quadratically to the statistical error derived in the source detection. The classification was then performed for a second time, using the individual 3$\sigma$ error circles as the search radius for each source. Figure \ref{fig:Phoenix} shows the combined EPIC image in the 0.2-4.5~keV energy range with H\textsc{i} contours from \citet{Young07} superimposed. Also shown in this figure is the combined \emph{RVB} band optical image from the Local Group Galaxy Survey (LGGS) image \citep{LGGS}. The positions of the sources detected by the analysis are also displayed (the boresight corrected positions on the optical image, the uncorrected positions on the X-ray image) along with a unique identification number for each source.

In order to locate any candidate HMXBs in Phoenix, it is necessary to identify as many of the foreground stars and background sources in our field of view as possible. Following the method of P04 we attempt to identify or classify all the X-ray sources in the field of view as either foreground stars (fg stars),\ AGN,\ galaxies (GAL),\ Super-soft Sources (SSS), Supernova Remnants (SNR) or hard sources (which could be HMXBs or unidentified AGN).

P04 use the hardness ratios (HRs) of the sources (both known and unknown) in M33 to create X-ray colour plots and identify areas of specific source class. The limitations of this method of source classification are discussed in detail in P04, the main drawback being that sources with similar spectra (e.g. foreground stars and supernova remnants; HMXBs and AGN) cannot be differentiated when the statistics are low. An alternative to HR classification is the quantile analysis technique of \citet{Hong04}, however this is not suitable for \emph{XMM-Newton} data due to the high background. The HRs are defined as $HR_i=(B_{i+1}-B_i)/(B_{i+1}+B_i)$ where $B_i$ is the count rate in energy band $i$. The bands are the same as those used in the source detection: (0.2--0.5), (0.5--1.0), (1.0--2.0), (2.0--4.5) and (4.5--12.0) keV. To improve the statistics, the counts from all three EPIC instruments were combined before HRs are calculated. The hydrogen column density in the direction of M33 is a factor of 4 greater than towards Phoenix ($1.5\times10^{21}$~cm$^{-2}$ for Phoenix and $6\times10^{21}$~cm$^{-2}$ for M33, \citealt{Dickey90}). Phoenix does not have the considerable X-ray coverage of M33 and so we did not attempt to derive our own HR criteria, instead we simulated model spectra to confirm that the HR criteria derived by P04 are valid for this work despite the difference in their column densities. Modifications were made where appropriate. 

\begin{figure}
\begin{center}
\includegraphics[height=170mm]{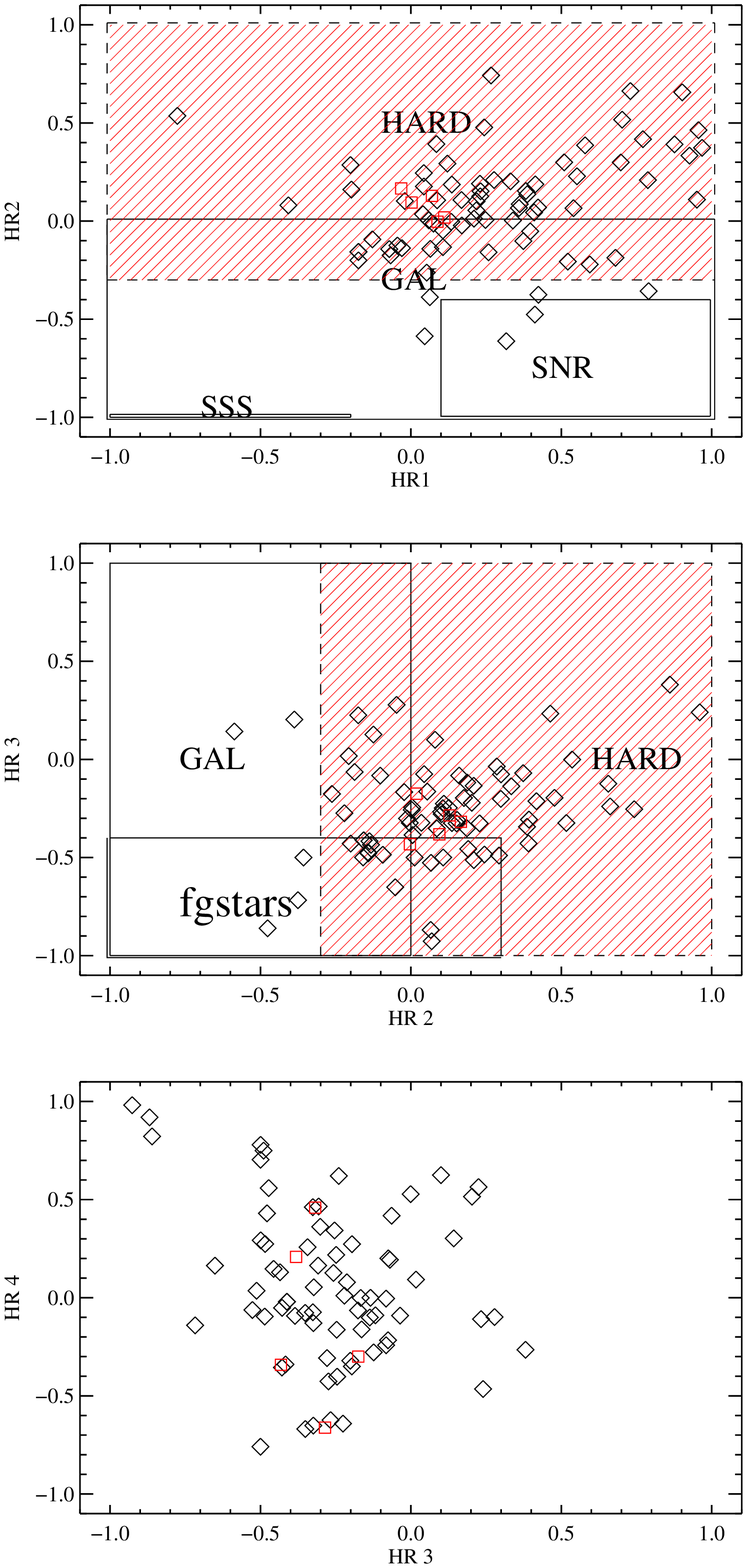}
\end{center}
\caption{X-ray hardness ratios for all sources in the field of view. The regions marked on the plots show the hardness ratio criteria for the different objects. We stress that for objects such as galaxies and foreground stars, it is the information gained at other wavelengths that drives the classification. The errors are not shown for clarity. The red squares are the five sources discussed in Section 5.}\label{fig:HRs}
\end{figure}

The modified and original P04 criteria used for source classification are summarised in Table \ref{table:criteria}. Where only one set of criteria are presented, no changes have been made to the original criteria. Where two sets of criteria are presented, the original criteria from P04 are listed underneath the modified criteria in italics. Figure \ref{fig:HRs} shows the X-ray hardness ratio plots for all the X-ray sources in Phoenix.

For sources such as foreground stars, galaxies and AGN, it is the additional information at other wavelengths that drives the classification and not the X-ray HRs. Foreground stars are classified based on the ratio of X-ray to optical flux, calculated using their magnitudes given by log$(\frac{f_X}{f_{opt}})=log(f_X)+(m_{V}/2.5)+5.37$ following \citet{Macca88}. If \emph{V}-band magnitude information was not available, the average of the $m_{B}$ and $m_{R}$ values were used in place of the $m_{V}$. The distances to all but a handful of stars in our own Galaxy are unknown, making it difficult to assess the level of extinction for an individual foreground star, regardless of the direction we are pointing. As such, no modifications were made to the HR criteria for foreground stars. 

\begin{table*}
\caption{Summary of criteria, identifications and classifications, for more details see text. EHR2 is the error on HR2.}\label{table:criteria}
\centering
\begin{tabular}{lcrr}
\hline\hline
Source Type & Selection Criteria & Identified & Classified \\
\hline
fg star & log($\frac{f_X}{f_{opt}}$)$<$ -1.0 and HR2$<$0.3 and HR3$<$-0.4 or not defined & 5 & 1 \\
SNR & HR1$>$0.1 and HR2$<$-0.4 and not a fg star & & \\
AGN & Radio source and classified hard & & 1 \\
GAL & optical id with a galaxy and HR2$<$0.0 & 1 & 3 \\
SSS & HR1$<$-0.2, HR2-EHR2$<$-0.99 or HR2 not defined, HR3 and HR4 not defined & &  \\
hard & HR2-EHR2$>$-0.3 or only HR3 and HR4 defined and no other classification &  & 50 \\
 & \emph{HR2-EHR2$>$-0.2 or only HR3 and HR4 defined and no other classification} & & \emph{48}\\
\hline
\end{tabular}
\end{table*}

The HR2\textless0 criterion for a galaxy classification is a constraint applied when an association was made between a galaxy and one of our X-ray sources. It is primarily to distinguish between a galaxy without an active nucleus and an AGN with a much harder spectrum. There is a smooth transition between the two classifications depending on the relative contributions of star formation and the activity of the central black hole. 

The HR criteria are more important for hard sources, supernova remnants and super-soft sources, where no other information is used in the classification. Super-soft sources are generally accepted to be white dwarf systems burning hydrogen on their surface. Any super-soft sources in Phoenix will reside in the bottom left-hand corner of the top panel of the HR plots (see Figure \ref{fig:HRs}). HRs were calculated for typical super-soft sources (absorbed 30~eV and 50~eV blackbodies) over a range of column densities (10$^{19}$--10$^{21}$~cm$^{-2}$) to see how the difference in the column density to M33 and Phoenix affects the HR values. All the HRs calculated were well within the original HR limits and so the criteria were left unchanged. 

The spectra of supernova remnants are very similar to those of foreground stars. Following P04 we assumed that all foreground stars have been identified from the correlation with optical data leaving any remaining sources that fit the HR criteria as supernova remnant candidates. Again, HRs were calculated for typical supernova remnant spectra (absorbed thermal plasma models with temperatures of 0.5~keV and 1.0~keV) over the same range of column densities as before. All the HRs calculated were within the original criteria and so the criteria were not modified.

The hard classification is arguably the most important for this work. Sources classified as hard are believed to be either unidentified AGN or HMXBs, both of which have absorbed power law spectra. HRs were calculated for an absorbed power law spectrum over a range of column densities (10$^{19}$--10$^{21}$ cm$^{-2}$) and photon indices (0.6--3.0). Only the very steepest photon indices ($\geq$2.8) with low absorption ($\leq2\times10^{20}$ cm$^{-2}$) did not fit the hard criterion. Such spectra would imply an AGN with very low intrinsic absorption. The criterion for hard sources were adjusted so that all photon indices with galactic absorption or greater are classified hard.

\section{Source List}

The X-ray hardness ratios of all sources in the field of view are displayed in Figure \ref{fig:HRs}. The red squares are the 5 sources that are identified in section 5 for further analysis. The black diamonds are the other 76 sources in the field of view. The hardness ratio criteria are also displayed. Only two X-ray sources have a SIMBAD object within their error circle, five matches were found within the NED database, 35 matches in the LGGS catalogue and a further 3 sources in the USNO-B1 catalogue. The criteria used for source classification are summarised in Table \ref{table:criteria}. It is important to note that the source classifications suggested here should be regarded as the proposed nature of the source. The original criteria were derived using the overall properties of large sample populations and not individual sources, consequently some characteristics of subclasses and sources are not accounted for, such as the X-ray soft, narrow line Seyfert I galaxies, and rare events like stellar flares. A source is \emph{``identified''} if it meets all the criteria for a particular group and is \emph{``classified''} if it meets the majority of the criteria. Nineteen sources remain unclassified. The source classifications of all the sources in the field of view can be found in the Appendix.

\subsection{Foreground stars}

Foreground stars are expected to lie in the lower left corner of the HR3/HR2 plot (middle panel of Figure \ref{fig:HRs}). Of the 38 X-ray sources in the field of view with an LGGS or USNO counterpart, one source (\#14) was positively identified as a star from the SIMBAD database and another source (\#56) from the NOMAD catalogue \citep{nomad}. A further two sources (\#27 and \#99) were identified as stars based on the optical information available in the LGGS and USNO catalogues.

Sources \#78 and \#84 all have log($\frac{f_X}{f_{opt}}$) indicating a stellar classification, but only source \#78 fits the hardness ratio criteria within errors so is also identified as a foreground star. Source \#84's HR3 value is greater than the -0.4 required for classification. The optical counterpart is listed in the NOMAD catalogue as being a 14th magnitude star with colours consistent with being a late K/early M type star. The ratio of X-ray to optical flux is also consistent with a K or M type star. We have already stressed in the previous section that for foreground stars, it is the information at other wavelengths that drives the classification. Despite the fact that the HR3 value does not fit the criterion, we feel that the strong agreement of the X-ray to optical flux value and the optical colours warrants a stellar classification. As such source \#84 is classified as a foreground star.

\subsection{Galaxies and AGN}

The X-ray spectrum of a galaxy without an active nucleus is softer than that of a galaxy with an AGN, and so will reside in the lower and left hand side of the HR1/HR2 and HR2/HR3 plots in Figure \ref{fig:HRs}. Cross correlation with the SIMBAD database identified one source, \#39, with a known galaxy in our field of view just outside the H\textsc{i} cloud. This corresponds to SUMSSJ015027.5-442537 at z=0.13. Four more X-ray sources (\#10, \#31, \#43 and \#60) have NED objects in their error circles. \#10, \#31, and \#60 all fit the X-ray HR criteria for galaxies. \#43 is not a galaxy, but listed as an unclassified extragalactic candidate within NED, detected in the radio band. The X-ray source just fits the criteria for a hard source and as such, \#43 is classified as an AGN. 

\subsection{Super-soft Sources and Supernova Remnants}

The area of Figure \ref{fig:HRs} where the super-soft sources should reside (bottom left-hand corner of the top panel) is remarkably empty. We do not identify or classify any sources as super-soft sources in our field of view. Supernova remnants are expected to fall in the area of the same plot marked SNR. Only two sources fit these HR criteria. They are \#14 and \#27, two identified foreground stars. As we are confident in our assessment of these sources, we don't find evidence for any SNR in our data.

\subsection{Extended sources}

Five sources were found to have some measurable extent, these were \#36, \#41, \#81, \#91 and \#101. With the exception of source \#41, these sources are all found on the edge of the field of view and close to a pn chip border. It is more likely that the elongation of the off-axis point spread function (PSF) is not well accounted for at these locations. \#41 is in very close proximity to \#39, the galaxy identified in the SIMBAD database and discussed in section 4.2. It has an extent of 5.7$\pm$0.2 pixels, which corresponds to 23.4\arcsec on the sky, greater than the separation of \#41 and \#39 (around $\sim$19\arcsec). Along with \#39, \#41 appears to be embedded within a single extended source in the X-ray image. It is likely that these two sources are in fact part of one extended source.

\section{HMXB candidates}

\begin{table*}
\caption{Properties of the sources detected in the H\textsc{i} region. Fluxes are from the model fits in the 0.2--10~keV energy range. For \#68, this information isn't available and so the 0.2--12.0~keV flux from the source detection is reported. Ratio of X-ray to optical value has been calculated following the method of \citet{Macca88} using the flux values from the model fit where possible.}\label{tab:HI}
\centering
\begin{tabular}{cccccccc}
\hline\hline 
\multirow{2}{*}{ID}  &    RA    &    Dec    &   err$_{1\sigma}$ &  \multirow{2}{*}{HR1} & \multirow{2}{*}{HR2} & \multirow{2}{*}{HR3} & \multirow{2}{*}{HR4} \\
 &   J2000    &    J2000    &  \arcsec & & & & \\
\hline\noalign{\smallskip}
35 & 01:50:41.0 & -44:24:32.9 & 1.5 & 0.11$\pm$0.06 & 0.02$\pm$0.06 & -0.17$\pm$0.06 & -0.30$\pm$0.09 \\
68 & 01:50:45.2 & -44:30:34.7 & 1.6 & 0.0$\pm$0.1 & 0.17$\pm$0.09 & -0.3$\pm$0.1 & 0.46$\pm$0.09 \\
52 & 01:50:47.2 & -44:27:44.2 & 1.5 & 0.09$\pm$0.03 & 0.00$\pm$0.03 & -0.43$\pm$0.03 & -0.34$\pm$0.06 \\
61 & 01:50:51.8 & -44:29:38.4 & 1.5 & 0.07$\pm$0.05 & 0.13$\pm$0.04 & -0.29$\pm$0.05 & -0.7$\pm$0.1 \\
34 & 01:51:00.5 & -44:24:29.7 & 1.6 & 0.00$\pm$0.09 & 0.09$\pm$0.08 & -0.4$\pm$0.1 & 0.2$\pm$0.2 \\
\hline\hline \noalign{\smallskip}
& $N_H$ & \multirow{2}{*}{$\Gamma$} & Flux & \multirow{2}{*}{$\chi^{2}_{r}$/d.o.f} & \multirow{2}{*}{$m_V$} & \multirow{2}{*}{\emph{B-V}} &  \multirow{2}{*}{log($\frac{f_X}{f_{opt}}$)} \\
 & $10^{20}~cm^{-2}$ & & ergs~cm$^{-2}$~s$^{-1}$ & & & \\
\hline\noalign{\smallskip}
35 & 0$^{+82}$ & 1.6$_{-0.2}^{+0.3}$ & 1.9$\times10^{-14}$ & 0.84/15 & 19.93$\pm$0.01 & 0.44$\pm$0.01 & -0.38 \\{\smallskip}
68 & & & 6.3$\pm$0.7$\times10^{-14}$ & & 21.22$\pm$0.02 & 0.28$\pm$0.02 & -0.12 \\{\smallskip}
52 & 5.2$_{-3.7}^{+5.2}$ & 2.1$\pm$0.1 & 6.5$\times10^{-14}$  & 1.21/54 & 19.49$\pm$0.01 & 0.21$\pm$0.01 & -0.14 \\{\smallskip}
61 & 5.2$_{-5.2}^{12.6}$ & 1.8$\pm$0.2 & 3.8$\times10^{-14}$ & 0.63/28 & 21.63$\pm$0.02 & 0.51$\pm$0.04 & 0.42 \\{\smallskip}
34 & 0$^{+13}$ & 1.9$_{-0.3}^{+0.4}$ & 1.2$\times10^{-14}$ & 1.58/16 & 21.47$\pm$0.02 & 0.23$\pm$0.04 & -0.12 \\
\hline
\end{tabular}
\label{table:HI}
\end{table*}

Of the 81 sources in the field of view, only one source, \#48, is coincident with the optical emission of the galaxy. Interestingly some of the brightest sources in the field of view are offset from the optical emission and spatially coincident with the H\textsc{i} region. Eleven sources (\#34, \#35, \#48, \#52, \#54, \#56, \#58, \#61, \#62, \#66 and \#68) lie within the lowest H\textsc{i} contour shown in Figure \ref{fig:Phoenix}. The recent bursts of star formation in the western side of the galaxy means that the location of these sources is compatible with a possible HMXB population. Source \#56 is the only identified source in this list. Five sources (\#34, \#35, \#52, \#61 and \#68) have appreciable X-ray emission above 2~keV (with flux values $>8\times10^{-15}$ ergs~cm$^{-2}$~s$^{-1}$) and occupy an area of \textless40 square arcminutes. Their HRs are shown by red squares in Figure \ref{fig:HRs}. Synthesis models of the cosmic X-ray background by \citet{Treister06} and \citet{Gilli07} suggest we should see $\sim2.3^{+1.2}_{-1.9}$ background sources in this energy range and in an area this size. The errors stated are the 90\% confidence limits. Whilst five is not an obvious excess over the predicted background numbers, we feel these sources warrant further investigation, particularly as the expected number of HMXBs in Phoenix is small.

Temporal and further spectral analysis are required to distinguish between HMXBs and as yet unidentified AGN. No other X-ray data are available for Phoenix, so we cannot draw comparisons between our results and past observations, but power spectra created from the entire (non GTI filtered) light curves for all five of these sources showed no evidence for pulsations.One thousand fake pulsar lightcurves were generated by adding a sinusoid with a period of 100~s to a constant equal to the background subtracted count rate of the brightest source. The light curves have an amplitude corresponding to a pulsed fraction of 0.3 (the typical value quoted for HMXBs in the SMC, \citealt{Coe10}). Random noise was added to every point in the light curve using the IDL program \textsf{poidev}, ensuring that the total number of counts in the light curve was conserved. The simulated light curves were then binned with the same bin time as the real light curves. When Lomb-Scargle analysis was performed, only 46 of the 1000 light curves resulted in a significant detection (a 5$\sigma$ detection according to the formula of \citealt{Horne86}). An unrealistically high  pulsed fraction of 0.6 is required to consistently produce a significant result at 100s with our total number of counts. Thus we can conclude that we do not have the sensitivity to detect pulsations from a typical pulsar in Phoenix using this method.

EPIC spectra were extracted between 200~eV and 10~keV for these sources and were fit with an absorbed powerlaw (\emph{phabs*vphabs*powerlaw}) model. The \emph{phabs} component is fixed at 1.5$\times10^{20}$~cm$^{-2}$ \citep{Dickey90} for galactic foreground absorption with elemental abundances from \citet{Wilms00}. The \emph{vphabs} component accounts for absorption in Phoenix and is a free parameter with metal abundances reduced to 0.075, calculated from values given in H09. Throughout the investigation XSPEC version 12.6.0 was used. Table \ref{tab:HI} contains the best fit parameters with 90\% confidence errors.

\citet{McB08} note that the spectral distribution of Be/X-ray binaries in the SMC is consistent with that of the Milky Way, despite the very different environment. Thus it seems reasonable to assume that any Be/X-ray binary in Phoenix will fall in the same range of spectral classes (O9V-B2V). Taking the $M_V$ magnitudes from the stellar flux library of \citet{Pickles98} and scaling them with the distance to Phoenix predicts \emph{V}-band magnitudes in the range 18.7-21.2. These values include an $A_V$ extinction value of 0.062~mags (H09). The distance modulus to Phoenix, 23.1, has an error of $\pm$0.1 (H09) which leads to the same error on the magnitudes calculated for Be/X-ray binaries in Phoenix. Similarly, if we assume any supergiant X-ray binary present will also fall into the same range of spectral classes as those already discovered in our own galaxy and beyond (O8.5I-B3I; \citealt{Liu06}) and scale to the distance of Phoenix, we predict \emph{V}-band magnitudes in the range 16.1-16.7 $\pm$0.1. When searching for HMXB candidates in the SMC, \citet{Shtyk05} required that the optical colours of the companion star were $B-V$\textless0.20. We adopt the same criteria for this work, but take into account the different levels of extinction between the SMC and Phoenix \citep{Rieke85}. This leads to the constraint that $B-V$\textless0.14 for any optical counterparts. 

Black hole X-ray binaries can exist in several states, the high/soft state, the low/hard state and the quiescent state (which can be considered to be a special case of the low/hard state; \citealt{Kong02}). The mass of Phoenix is too low for an LMXB population, but no such restriction applies to the high mass black hole systems. Our observations are not deep enough to see any quiescent black hole binaries in Phoenix, which would have a flux of about $\sim5\times10^{-17}$~ergs~cm$^{-2}$ ~s$^{-1}$\citep{McC06}. The luminosity of the high/soft state is typically seen around 10\% of the Eddington luminosity \citep{Nowak95}, for a 6M$_\odot$ black hole at the distance of Phoenix, this is $\sim8\times10^{37}$~ergs~s$^{-1}$ corresponding to a flux of $\sim4\times10^{-12}$~ergs~cm$^{-2}$ ~s$^{-1}$. This is an order of magnitude brighter than even the brightest source in our field of view. Consquently, any black hole binaries present in our data set must be in the low/hard state, which is charactised by a power law spectrum with a photon index $\sim$1.7 and a luminosity around $\sim$2\% Eddington or less \citep{McC06,Maccarone03}. In the X-ray band, Be/X-ray binaries typically have photon indices of $\leq$1.4 \citep{Haberl08}.

\subsection{Source \#34}

Source \#34 is one of the two spectra with lower signal and as such the parameters for the model fit are not well constrained. The photon index of 1.9$_{-0.3}^{+0.4}$ rules out the possibility that \#34 is a Be/X-ray binary but is consistent with a black hole X-ray binary in Phoenix. The large uncertainty in the parameters and the relatively poor fit means that this should be treated with caution. The optical counterpart is too faint and red for an early type star ($m_V$=21.5, ${B-V}$=0.23). Source \#34 is most likely an AGN.

\subsection{Source \#35}

The X-ray spectrum of source \#35 suffers from the same low signal problems as \#34. The photon index (1.6$_{-0.2}^{+0.3}$) is just consistent with that of a neutron star in a Be/X-ray binary but comfortably within the range of indices seen for black hole binaries and AGN. The optical counterpart has $m_V$ consistent with a B-type star in Phoenix, but is well outside the acceptable colour range. as with \#34, \#35 is probably an AGN.

\subsection{Source \#52}

The photon index of \#52 (2.1$\pm$0.1) is too soft for even a black hole X-ray binary, and so source \#52 is almost certainly an AGN. The optical information for this source supports this as again, it is far too red for an early type star in Phoenix (${B-V}$=0.21$\pm$0.01).

\begin{figure*}
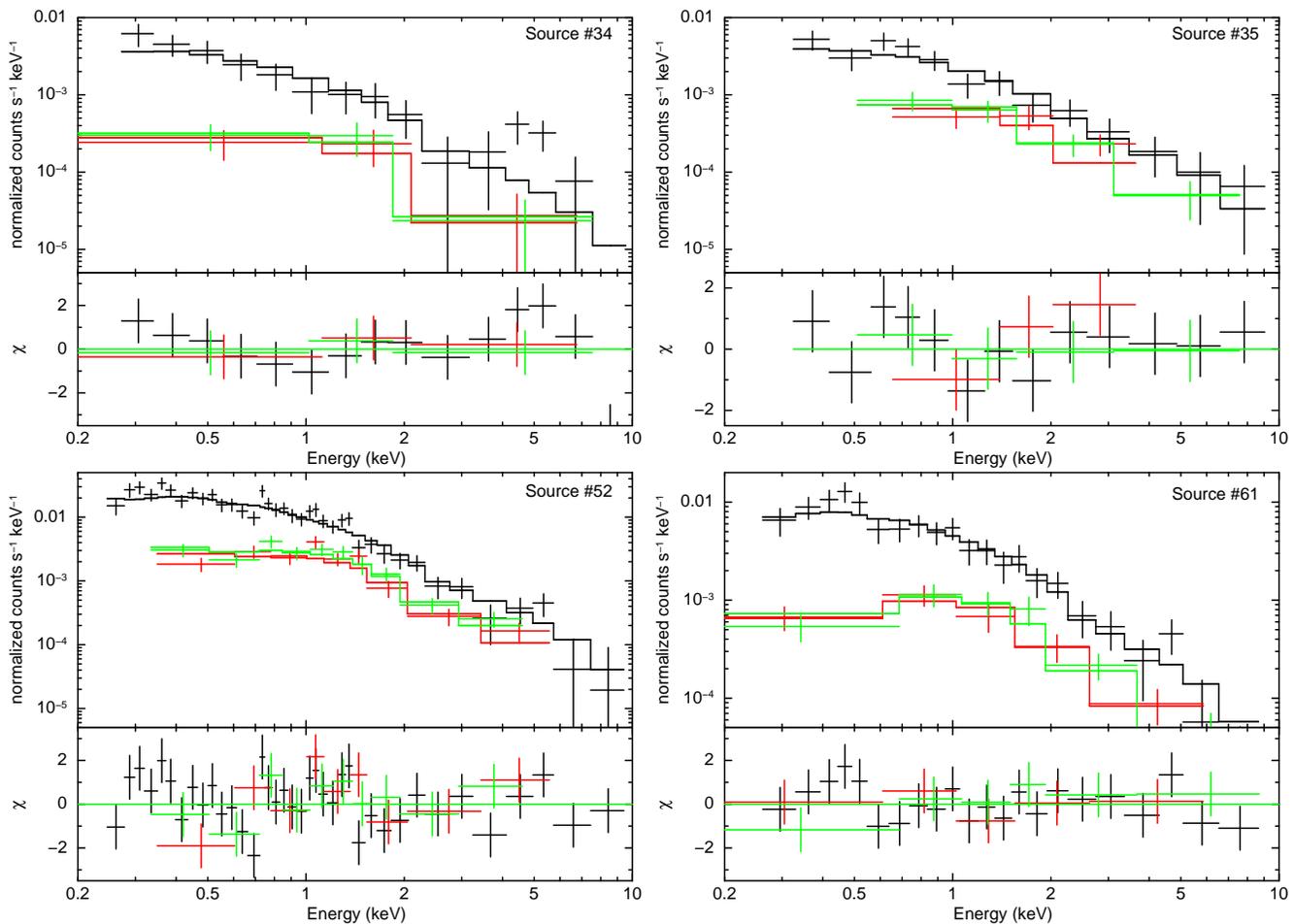

  \begin{center}
  \includegraphics[height=0.5\textwidth, angle=270]{spectrum_34x2.ps}\includegraphics[height=0.5\textwidth, angle=270]{spectrum_35x2.ps}
  \includegraphics[height=0.5\textwidth, angle=270]{spectrum_52x2.ps}\includegraphics[height=0.5\textwidth, angle=270]{spectrum_61x2.ps}
  \caption{X-ray spectra of four of the sources in the H\textsc{i} region and their best fit models.}\label{fig:spec}
  \end{center}
\end{figure*}

\subsection{Source \#61}

Source \#61 has the greatest X-ray to optical luminosity ratio of our sample and the second greatest in the entire field of view (source \#28 has a slightly greater log($\frac{f_X}{f_{opt}}$) value, but is located further from the optical centre of Phoenix and on the eastern side of the galaxy). The photon index (1.8$\pm$0.2) is consistent with a black hole X-ray binary in Phoenix (though also consistent with an AGN). Its optical counterpart is too faint and red for an early type star ($m_V$=21.6, ${B-V}$=0.51). Source \#61 is again probably an AGN.

\subsection{Source \#68}

Source \#68 is an unfortunately placed X-ray source, falling on a bad column on the EPIC-pn detector and near a chip border in both MOS detectors. The total number of counts for this source was not sufficient to extract a spectrum and so few conclusions can be drawn. The optical counterpart has brightness, but not colours, consistent with an HMXB in Phoenix ($m_V$=21.2, ${B-V}$=0.28) and the 0.2--12.0 keV flux value derived from the source detection (6.3$\times10^{-14}$~ergs~cm$^{-2}$~s$^{-1}$) suggests a luminosity of 1$\times10^{36}$~ergs s$^{-1}$ (assuming \#68 is associated with Phoenix). However this flux value should be interpreted with care as detector edge effects may introduce an additional uncertainty that cannot be accounted for. Source \#68 is also probably an AGN.

\section{Other sources associated with Phoenix}

Interestingly, only one source in our field, \#48, is coincident with the most obvious optical emission of the Phoenix dwarf galaxy. Massive black holes have been discovered in nearby dwarf galaxies (for example see \citealt{Wassenhove10}) and the location of the source, close to what appears to be the centre of the galaxy, led us to consider the source as a possible massive black hole candidate. Although radial velocities have been measured for a handful of stars in Phoenix \citep{Gallart01,Irwin02} no information on the location of the dynamical centre is available.

An estimate of the position of the galaxy centre has been obtained by maximising the number of LGGS sources within a circular region when the centre and size of the region is varied, following the method of \citet{Dieball2010}. The uncertainties are estimated via a bootstrapping method. One thousand artificial catalogues were created by sampling with replacement from the original LGGS catalogue. The centre was then estimated for each of these artificial catalogues using the same method detailed above. The error on our position is taken to be the standard deviation of these centre estimates. Our final result was obtained using a circular region with a 150 pixel (40.5\arcsec) radius, but experiments with other region sizes produced consistent results. The galaxy centre was put at RA=01\textsuperscript{h}51\textsuperscript{m}8.2\textsuperscript{s}$\pm$6.1\arcsec, Dec=-44$^{\circ}$26\arcmin55.9\arcsec$\pm$7.6\arcsec. This position should be treated with caution as the LGGS catalogue is by no means complete.

Source \#48 is situated 40\arcsec away from our centre position with a position error of 1.7\arcsec. This mis-match in positions suggests that source \#48 is unlikely to be a massive black hole located at the centre of the Phoenix galaxy. The source hardness ratios are also not consistent with a black hole/AGN in Phoenix. However, the errors on all these values are large and better X-ray and optical data are needed before drawing any firm conclusions about this source.

\section{Discussion}

Of the 81 X-ray sources detected in the \emph{XMM-Newton} field of view, we have categorised six of the sources as foreground stars, four as background galaxies and one as a possible AGN. The majority of the remaining sources in the field of view with optical counterparts have log($\frac{f_X}{f_{opt}}$) values between -1 and 1, typical of local AGNs and high star forming galaxies \citep{Hornschemeier01}. Source \#96 has the greatest log($\frac{f_X}{f_{opt}}$) value at 1.13, putting it at the very X-ray brightest end of these objects and a possible BL Lac candidate. Further analysis of 5 potential HMXB candidates in the field of view suggests they are all AGN.

We find no evidence for any black hole binaries in Phoenix. The mass of Phoenix is too low for us to expect any LMXBs, so any black hole binaries present would be expected to be high mass black hole X-ray binaries, in the low/hard state. \citet{Belczynski09} use binary population synthesis models to show that the expected ratio of Be/X-ray binaries with neutron stars to black holes in the Galaxy is relatively high ($\sim$30). Thus far, not a single Be/X-ray binary has been found to host a black hole. This leaves only black hole supergiant systems. These systems are a small fraction of the total HMXB population in our own Galaxy and are not found at all in the SMC (e.g. \citealt{Coe09}). If the population of HMXBs is as small as predicted, the presence of a black hole supergiant HMXB in Phoenix seems unlikely.

If these sources are spectroscopically confirmed to be AGN, their position behind the gas cloud and their magnitude will make them an excellent tool to further probe the metallicity of this region using the Na\textsc{i}D and Ca\textsc{ii}K lines (for e.g. see \citealt{VanLoon09}). Accurate measurements of the Na and Ca column density across these sources could tell us more about the gas cloud and may even reveal inhomogeneities in the chemical composition across the region.

\citet{Ant10} report that Be/X-ray binaries in the SMC are found in regions where a burst of star formation occurred $\sim$25-60~Myr ago. As such, the five X-ray sources identified offset to the west main stellar body where a strong episode of star formation took place around $\sim$40~Myr ago were good candidates for Be/X-ray binaries. However, none of these sources display all the X-ray and optical characteristics of a Be/X-ray binary. The strongest evidence for a neutron star is periodic modulation of the X-ray flux. We do not have the required sensitivity to detect this in the data but this condition is sufficient, not necessary, for proof of a neutron star's existence.

In general, Be/X-ray binaries are by far the most numerous subclass of HMXBs \citep{Liu06} and so have the greatest likelihood of existence in Phoenix. However, as supergiant systems hosting neutron stars are persistent sources of X-rays, we may have a better chance of seeing them in a single epoch. The lack of evidence for any supergiant binaries, with either a black hole or a neutron star secondary, would strongly suggest that they do not exist in Phoenix. This is in line with current predictions based on star formation rate \citep{Grimm03} and is further evidence in support of this relation. Unlike supergiant systems, Be/X-ray binaries are transient in nature and so there is a possibility that a greater population exists than is hinted at here. From our regular observations of the SMC over a 10~yr interval, we find that Be/X-ray binaries are in outburst for around $\sim$10\% of an orbital period and are active, on average, for $\sim$20\% of the time \citep{Galache08,McG08}. It follows that the probability of seeing a particular Be/X-ray binary active at any one moment is about $\sim$2\%. Should the number of Be/X-ray binaries in Phoenix be limited to just 1--2 systems, then the probability of seeing just one system active at any moment in time is \textless10\%. Further observations of the Phoenix dwarf galaxy could yet reveal a HMXB population.

This null result is consistent with extending the LogN-LogS relationship reported by \citet{Grimm03} for HMXBs down to lower luminosities and lower star formation rates. Though a positive number would have provided more substantial agreement, it is now evident that a large population of low luminosity sources does not exist - at least not in Phoenix.

\section*{Acknowledgments}

Based on observations obtained with XMM-Newton, an ESA science mission with instruments and contributions directly funded by ESA Member States and NASA. This research has made use of the SIMBAD database, operated at CDS, Strasbourg, France and the NASA/IPAC Extragalactic Database (NED) which is operated by the Jet Propulsion Laboratory, California Institute of Technology, under contract with the National Aeronautics and Space Administration.  ESB acknowledges support from a Science and Technology Facilities Council studentship.

\label{lastpage}
\bsp
\appendix
\begin{landscape}
\begin{table}
\section{Complete X-ray Source List}
\caption{Properties of all the sources in the \emph{XMM-Newton} field of view. Count rates and fluxes are both given for the 0.2--12~keV energy range. The log($\frac{f_X}{f_{opt}}$) is calculated from the 0.2--4.5 keV flux values and the optical data available in the LGGS or USNO catalogues. The classification column follows the designation in the text with \emph{classified} sources in square brackets. Unclassified sources are listed as ``unclass''.}\label{tab:app}
\centering
\begin{tabular}{cccccccccccc}
\hline\hline 
\multirow{2}{*}{ID}  &    RA    &    Dec    &   err$_{1\sigma}$    &  Count rate & flux & \multirow{2}{*}{log($\frac{f_X}{f_{opt}}$)} & \multirow{2}{*}{HR1} & \multirow{2}{*}{HR2} & \multirow{2}{*}{HR3} & \multirow{2}{*}{HR4} & \multirow{2}{*}{Classification}\\
 &    J2000    &    J2000    &   \arcsec    &    10$^{-3}$~s$^{-1}$    &  10$^{-14}$ ergs~cm$^{-2}$~s$^{-1}$ &  &  & &  & &\\
\hline\noalign{\smallskip}
47 & 01:49:43.7 & -44:26:46.4 & 2.1 & 10$\pm$1 & 2.4$\pm$0.8 & -0.05 & 0.0$\pm$0.1	&-0.1$\pm$0.1&0.1$\pm$0.2&-1.0$\pm$0.3 & [hard]\\
23 & 01:49:54.1 & -44:22:49.3 & 1.7 & 12$\pm$2 & 8$\pm$1 &  & 0.7$\pm$0.3&0.7$\pm$0.1&-0.2$\pm$0.1&	0.6$\pm$0.1&[hard]\\
36 & 01:49:54.4 & -44:25:04.0 & 1.7 & 11$\pm$1 & 8$\pm$1 &  & 0.4$\pm$0.2&0.1$\pm$0.1&-0.9$\pm$	0.2&0.98$\pm$0.07&unclass\\
81 & 01:49:54.8 & -44:32:27.9 & 1.6 & 24$\pm$2 & 12$\pm$2 &  & -0.41$\pm$0.08&0.1$\pm$0.1&0.1$\pm$0.1&0.62$\pm$	0.06&[hard]\\
91 & 01:50:02.1 & -44:34:56.2 & 1.8 & 17$\pm$2 & 6$\pm$2 &  & -0.78$\pm$0.09&0.5$\pm$0.1&0.0$\pm$0.1&0.53$\pm$0.07&[hard]\\
15 & 01:50:03.5 & -44:21:23.5 & 1.7 & 5.9$\pm$0.8 & 0.6$\pm$0.3 &  & 0.2$\pm$0.1&0.1$\pm$0.1&-0.5$\pm$0.2&0.8	$\pm$	0.3	&unclass\\
84 & 01:50:06.8 & -44:33:20.0 & 1.8 & 11$\pm$1 & 1.0$\pm$0.9 & -3.10 & 0.05	$\pm$	0.09	&	-0.6	$\pm$	0.1	&	0.1	$\pm$	0.3	&	0.3	$\pm$	0.3&[fgstar]\\
25 & 01:50:07.6 & -44:22:54.4 & 1.6 & 24$\pm$1& 3.9$\pm$0.7 & -0.02 & 0.04	$\pm$	0.06	&	0.04	$\pm$	0.06	&	-0.32	$\pm$	0.08	&	0.1	$\pm$	0.1&[hard]\\
31 & 01:50:11.5 & -44:23:23.2 & 1.5 & 30$\pm$1 & 5.3$\pm$0.7 & 0.18 & 0.11	$\pm$	0.05	&	-0.13	$\pm$	0.05	&	-0.43	$\pm$	0.07	&	0.1	$\pm$	0.1&[gal]\\
49 & 01:50:12.7 & -44:26:58.6 & 1.5 & 33$\pm$1 & 11$\pm$1 & -0.30 & 0.70	$\pm$	0.07	&	0.52	$\pm$	0.04	&	-0.32	$\pm$	0.04	&	-0.13	$\pm$	0.0&[hard]\\
33 & 01:50:14.9 & -44:24:04.2 & 1.6 & 7.0$\pm$0.8 & 1.4$\pm$0.4 &  & 0.9	$\pm$	0.2	&	0.66	$\pm$	0.09	&	-0.1	$\pm$	0.1	&	-0.3	$\pm$	0.2&[hard]\\
70 & 01:50:15.0 & -44:30:41.1 & 1.6 & 8$\pm$1 & 2.2$\pm$0.7 &  & 0.1	$\pm$	0.1	&	-0.1	$\pm$	0.2	&	-0.5	$\pm$	0.1	&	0.4	$\pm$	0.2	&unclass\\
96 & 01:50:15.4 & -44:35:51.4 & 1.5 & 67$\pm$2& 19$\pm$2 & 1.13 & 0.23	$\pm$	0.04	&	0.15	$\pm$	0.04	&	-0.33	$\pm$	0.04	&	-0.07	$\pm$	0.08	&[hard]\\
7 & 01:50:19.4 & -44:39:56.0 & 1.5 & 58$\pm$2 & 13$\pm$2 & -0.44 & -0.17	$\pm$	0.04	&	-0.20	$\pm$	0.05	&	-0.43	$\pm$	0.07	&	-0.4	$\pm$	0.3&unclass\\
71 & 01:50:19.7 & -44:30:47.7 & 1.6 & 19$\pm$1& 3.4$\pm$0.5 & 0.87 & 0.04	$\pm$	0.08	&	0.18	$\pm$	0.07	&	-0.20	$\pm$	0.07	&	-0.4	$\pm$	0.1&[hard]\\
3 & 01:50:20.5 & -44:17:10.6 & 1.7 & 11$\pm$1 & 6$\pm$1 &  & 0.1	$\pm$	0.2	&	0.4	$\pm$	0.1	&	-0.3	$\pm$	0.1	&	0.5	$\pm$	0.1	&[hard]\\
42 & 01:50:20.7 & -44:25:55.5 & 1.7 & 6.9$\pm$0.8 & 3.0$\pm$0.7 & 0.08 & 0.2	$\pm$	0.1	&	0.0	$\pm$	0.1	&	-0.5	$\pm$	0.2	&	0.7	$\pm$	0.1&unclass\\
4 & 01:50:23.8 & -44:17:08.7 & 1.6 & 17$\pm$2& 2.2$\pm$0.6 & 0.15 & 0.1	$\pm$	0.1	&	0.0	$\pm$	0.1	&	-0.3	$\pm$	0.1	&	0.4	$\pm$	0.1	&[hard]\\
14 & 01:50:24.2 & -44:20:49.1 & 1.6 & 6.8$\pm$0.7 & 0.9$\pm$0.4 & -4.55 & 0.32	$\pm$	0.08	&	-0.61	$\pm$	0.09	&	-1.0	$\pm$	0.4	&	1.0	$\pm$	1.9&fgstar\\
39 & 01:50:28.1 & -44:25:35.2 & 1.7 & 11.6$\pm$0.7 & 1.5$\pm$0.3 & -0.90 & 0.79	$\pm$	0.05	&	-0.36	$\pm$	0.06	&	-0.5	$\pm$	0.1	&	-0.8	$\pm$	0.3&gal\\
11 & 01:50:28.7 & -44:19:27.7 & 1.7 & 8.6$\pm$0.8 & 5.8$\pm$0.9 &  & 0.3	$\pm$	0.3	&	0.7	$\pm$	0.1	&	-0.3	$\pm$	0.1	&	0.3	$\pm$	0.1&[hard]\\
41 & 01:50:29.0 & -44:25:51.7 & 1.6 & 15.3$\pm$0.8 & 3.0$\pm$0.4 &  & 0.40	$\pm$	0.06	&	-0.05	$\pm$	0.06	&	-0.65	$\pm$	0.07	&	0.2	$\pm$	0.1&[gal]\\
82 & 01:50:32.9 & -44:32:32.2 & 1.6 & 9.4$\pm$0.8 & 1.4$\pm$0.4 & 0.27 & 0.2	$\pm$	0.1	&	0.10	$\pm$	0.09	&	-0.3	$\pm$	0.1	&	-0.6	$\pm$	0.3&[hard]\\
90 & 01:50:38.5 & -44:34:45.5 & 1.5 & 60$\pm$2& 9.2$\pm$0.8 & 0.27 & -0.03	$\pm$	0.03	&	-0.14	$\pm$	0.04	&	-0.42	$\pm$	0.05	&	-0.3	$\pm$	0.1&unclass\\
1 & 01:50:39.2 & -44:40:25.5 & 1.9 & 9$\pm$2 & 5$\pm$2 &  & -0.2	$\pm$	0.2	&	0.3	$\pm$	0.2	&	0.0	$\pm$	0.2	&	-0.1	$\pm$	0.4	&[hard]\\
56 & 01:50:42.7 & -44:29:04.5 & 1.5 & 22.4$\pm$0.8 & 2.5$\pm$0.2 & -2.20 & 0.42	$\pm$	0.04	&	-0.38	$\pm$	0.04	&	-0.72	$\pm$	0.06	&	-0.1	$\pm$	0.2 &fgstar\\
99 & 01:50:43.6 & -44:36:13.1 & 1.6 & 12$\pm$1 & 2.6$\pm$0.8 & -4.46 & 0.2	$\pm$	0.1	&	0.19	$\pm$	0.09	&	-0.5	$\pm$	0.1	&	0.1	$\pm$	0.2 &fgstar\\
54 & 01:50:45.3 & -44:28:29.2 & 1.6 & 7.1$\pm$0.5 & 1.5$\pm$0.3 & 0.08 & 0.26	$\pm$	0.09	&	-0.16	$\pm$	0.09	&	-0.5	$\pm$	0.1	&	0.3	$\pm$	0.2 &unclass\\
12 & 01:50:46.6 & -44:20:31.6 & 1.7 & 4.4$\pm$0.5 & 2.0$\pm$0.5 &  & 0.4	$\pm$	0.2	&	0.0	$\pm$	0.1	&	-0.1	$\pm$	0.2	&	0.2	$\pm$	0.2 &[hard]\\
100 & 01:50:54.5 & -44:36:23.7 & 1.7 & 9.4$\pm$0.9 & 2.3$\pm$0.6 & -0.31 & -0.1	$\pm$	0.1	&	-0.1	$\pm$	0.1	&	-0.5	$\pm$	0.2	&	0.6	$\pm$	0.2 &unclass\\
62 & 01:50:54.6 & -44:29:44.7 & 1.7 & 3.4$\pm$0.4 & 0.9$\pm$0.3 & -0.07 & 0.1	$\pm$	0.2	&	0.0	$\pm$	0.2	&	0.3	$\pm$	0.2	&	-0.1	$\pm$	0.2 &[hard]\\
66 & 01:51:00.1 & -44:30:19.6 & 1.7 & 13$\pm$1 & 1.4$\pm$0.3 &  & 0.4	$\pm$	0.1	&	0.2	$\pm$	0.1	&	-0.3	$\pm$	0.1	&	0.2	$\pm$	0.1	&[hard]\\
9 & 01:51:02.9 & -44:18:52.4 & 1.5 & 21$\pm$1 & 9.4$\pm$0.8 &  & 0.77	$\pm$	0.09	&	0.42	$\pm$	0.06	&	-0.21	$\pm$	0.06	&	0.08	$\pm$	0.07	&[hard]\\
16 & 01:51:03.2 & -44:21:39.3 & 1.8 & 3.5$\pm$0.4 & 1.0$\pm$0.3 &  & 0.3	$\pm$	0.2	&	0.2	$\pm$	0.1	&	-0.2	$\pm$	0.1	&	0.0	$\pm$	0.2&[hard]\\
58 & 01:51:04.3 & -44:29:23.7 & 1.6 & 9.9$\pm$0.6 & 1.5$\pm$0.2 & 0.07 & 0.05	$\pm$	0.06	&	-0.26	$\pm$	0.07	&	-0.2	$\pm$	0.1	&	-0.1	$\pm$	0.1&unclass\\
48 & 01:51:04.6 & -44:26:49.2 & 1.7 & 2.8$\pm$0.3 & 0.4$\pm$0.2 & -0.61 & -0.2	$\pm$	0.1	&	-0.2	$\pm$	0.2	&	-0.4	$\pm$	0.2	&	0.0	$\pm$	0.4&unclass\\
10 & 01:51:09.0 & -44:18:56.0 & 1.6 & 20.7$\pm$0.8 & 4.3$\pm$0.5 & 0.21 & -0.13	$\pm$	0.04	&	-0.09	$\pm$	0.05	&	-0.49	$\pm$	0.07	&	-0.1	$\pm$	0.2&[gal]\\
21 & 01:51:13.7 & -44:22:46.2 & 1.7 & 8.7$\pm$0.9 & 1.1$\pm$0.3 &  & 0.9	$\pm$	0.1	&	0.3	$\pm$	0.1	&	-0.1	$\pm$	0.1	&	-0.1	$\pm$	0.1&[hard]\\
28 & 01:51:13.8 & -44:23:07.8 & 1.6 & 19$\pm$1 & 3.2$\pm$0.4 & 0.47 & 0.22	$\pm$	0.08	&	0.05	$\pm$	0.08	&	-0.16	$\pm$	0.08	&	-0.2	$\pm$	0.1&[hard]\\
38 & 01:51:14.9 & -44:25:09.9 & 1.6 & 9.8$\pm$0.5 & 1.5$\pm$0.2 &  & 0.39	$\pm$	0.07	&	0.14	$\pm$	0.06	&	-0.32	$\pm$	0.07	&	-0.7	$\pm$	0.1&[hard]\\
2 & 01:51:17.6 & -44:40:14.9 & 1.6 & 9$\pm$1 & 2.3$\pm$0.8 &  & 0.1	$\pm$	0.1	&	-0.4	$\pm$	0.1	&	0.2	$\pm$	0.2	&	0.5	$\pm$	0.3&unclass\\
\hline
\end{tabular}
\end{table}
\end{landscape}
\begin{landscape}
\begin{table}
\centering
\begin{tabular}{cccccccccccc}
\hline\hline 
\multirow{2}{*}{ID}  &    RA    &    Dec    &   err$_{1\sigma}$    &  Count rate & flux & \multirow{2}{*}{log($\frac{f_X}{f_{opt}}$)} & \multirow{2}{*}{HR1} & \multirow{2}{*}{HR2} & \multirow{2}{*}{HR3} & \multirow{2}{*}{HR4} & Classification\\
 &    J2000    &    J2000    &   \arcsec    &    10$^{-3}$~s$^{-1}$    &  10$^{-14}$ ergs~cm$^{-2}$~s$^{-1}$ &   &  & &  & &\\
\hline\noalign{\smallskip}
22 & 01:51:21.6 & -44:22:43.2 & 1.6 & 16$\pm$1 & 1.3$\pm$0.2 &  & 0.04	$\pm$	0.09	&	0.24	$\pm$	0.08	&	-0.48	$\pm$	0.08	&	0.27	$\pm$	0.08	&unclass\\
55 & 01:51:22.5 & -44:28:39.6 & 1.6 & 7.3$\pm$0.5 & 1.1$\pm$0.2 &  & 0.14	$\pm$	0.09	&	0.19	$\pm$	0.08	&	-0.35	$\pm$	0.09	&	-0.7	$\pm$	0.2&[hard]\\
13 & 01:51:23.3 & -44:20:44.0 & 1.6 & 7.1$\pm$0.6 & 1.7$\pm$0.3 &  & 0.9	$\pm$	0.1	&	0.39	$\pm$	0.07	&	-0.43	$\pm$	0.08	&	-0.1	$\pm$	0.2&[hard]\\
92 & 01:51:24.7 & -44:35:09.2 & 1.6 & 21$\pm$1 & 10.0$\pm$0.9 & -0.82 & 1.0	$\pm$	0.6	&	0.86	$\pm$	0.07	&	0.38	$\pm$	0.05	&	-0.27	$\pm$	0.0&[hard]\\
50 & 01:51:24.9 & -44:27:23.9 & 1.6 & 6.8$\pm$0.6 & 1.6$\pm$0.3 &  & 0.2	$\pm$	0.1	&	0.0	$\pm$	0.1	&	-0.2	$\pm$	0.1	&	0.0	$\pm$	0.2&[hard]\\
72 & 01:51:25.9 & -44:31:07.4 & 1.7 & 10$\pm$1 & 2.4$\pm$0.5 &  & -0.1	$\pm$	0.2	&	-0.2	$\pm$	0.2	&	0.2	$\pm$	0.2	&	0.6	$\pm$	0.1	&unclass\\
29 & 01:51:26.0 & -44:23:11.9 & 1.7 & 4.8$\pm$0.5 & 1.0$\pm$0.2 &  & 0.4	$\pm$	0.1	&	0.1	$\pm$	0.1	&	-0.5	$\pm$	0.1	&	-0.1	$\pm$	0.2&unclass\\
20 & 01:51:28.6 & -44:22:39.1 & 1.9 & 2.9$\pm$0.4 & 0.3$\pm$0.1 & -0.84 & 1.0	$\pm$	0.2	&	0.1	$\pm$	0.1	&	-0.2	$\pm$	0.1	&	-0.6	$\pm$	0.3&[hard]\\
73 & 01:51:30.1 & -44:30:54.3 & 1.5 & 23.2$\pm$0.9 & 5.2$\pm$0.5 & 0.12 & 0.25	$\pm$	0.05	&	0.01	$\pm$	0.05	&	-0.39	$\pm$	0.06	&	-0.1	$\pm$	0.1&[hard]\\
65 & 01:51:30.2 & -44:29:57.0 & 1.6 & 17$\pm$2 & 3.0$\pm$0.5 &  & 0.4	$\pm$	0.1	&	0.19	$\pm$	0.09	&	-0.1	$\pm$	0.1	&	-0.1	$\pm$	0.1	&[hard]\\
27 & 01:51:30.6 & -44:22:54.9 & 1.6 & 9.0$\pm$0.5 & 1.2$\pm$0.2 & -1.41 & 0.41	$\pm$	0.06	&	-0.48	$\pm$	0.06	&	-0.9	$\pm$	0.1	&	0.8	$\pm$	0.2&fgstar\\
74 & 01:51:32.1 & -44:31:11.3 & 1.7 & 5.4$\pm$0.6 & 1.6$\pm$0.4 &  & 0.5	$\pm$	0.2	&	0.1	$\pm$	0.1	&	-0.9	$\pm$	0.1	&	0.9	$\pm$	0.1&unclass\\
59 & 01:51:37.3 & -44:29:27.3 & 1.6 & 7.6$\pm$0.6 & 1.4$\pm$0.3 &  & 0.36	$\pm$	0.09	&	0.09	$\pm$	0.08	&	-0.4	$\pm$	0.1	&	-0.1	$\pm$	0.2&[hard]\\
63 & 01:51:39.7 & -44:29:46.8 & 1.6 & 9.9$\pm$0.7 & 3.3$\pm$0.5 &  & 0.34	$\pm$	0.09	&	0.00	$\pm$	0.08	&	-0.3	$\pm$	0.1	&	0.1	$\pm$	0.1&[hard]\\
8 & 01:51:41.3 & -44:18:27.3 & 1.5 & 35$\pm$1 & 7.0$\pm$0.7 & -0.02 & 0.06	$\pm$	0.04	&	0.00	$\pm$	0.04	&	-0.25	$\pm$	0.05	&	-0.40	$\pm$	0.09&[hard]\\
6 & 01:51:41.4 & -44:18:13.3 & 1.7 & 6.5$\pm$0.9 & 2.6$\pm$0.6 &  & 0.1	$\pm$	0.2	&	0.3	$\pm$	0.2	&	-0.5	$\pm$	0.2	&	0.7	$\pm$	0.1	&unclass\\
67 & 01:51:41.8 & -44:30:13.6 & 1.7 & 7.2$\pm$0.6 & 2.3$\pm$0.4 &  & 0.5	$\pm$	0.1	&	-0.2	$\pm$	0.1	&	0.0	$\pm$	0.1	&	0.1	$\pm$	0.1	&unclass\\
85 & 01:51:42.0 & -44:33:21.7 & 1.6 & 11.9$\pm$0.9 & 2.9$\pm$0.7 &  & 0.3	$\pm$	0.1	&	0.21	$\pm$	0.07	&	-0.51	$\pm$	0.09	&	0.0	$\pm$	0.2	&unclass\\
45 & 01:51:42.7 & -44:26:07.7 & 1.6 & 13.0$\pm$0.7 & 3.1$\pm$0.4 &  & 0.23	$\pm$	0.07	&	0.13	$\pm$	0.06	&	-0.25	$\pm$	0.07	&	-0.2	$\pm$	0.1	&[hard]\\
17 & 01:51:43.3 & -44:21:55.9 & 1.9 & 4.6$\pm$0.6 & 1.8$\pm$0.5 &  & 1.0	$\pm$	0.1	&	0.4	$\pm$	0.1	&	-0.1	$\pm$	0.1	&	0.2	$\pm$	0.2	&[hard]\\
94 & 01:51:44.5 & -44:35:31.1 & 1.7 & 11$\pm$1 & 6$\pm$1 &  & 0.6	$\pm$	0.1	&	0.2	$\pm$	0.1	&	-0.3	$\pm$	0.1	&	0.5	$\pm$	0.2	&[hard]\\
18 & 01:51:45.6 & -44:22:00.0 & 1.6 & 13.1$\pm$0.8 & 2.3$\pm$0.4 & -0.58 & -0.02	$\pm$	0.07	&	0.10	$\pm$	0.07	&	-0.28	$\pm$	0.08	&	-0.3	$\pm$	0.2	&[hard]\\
101 & 01:51:49.9 & -44:37:04.5 & 1.7 & 12$\pm$1& 2.5$\pm$0.7 & -0.48 & -0.2	$\pm$	0.1	&	0.2	$\pm$	0.1	&	-0.1	$\pm$	0.1	&	0.0	$\pm$	0.2	&[hard]\\
78 & 01:51:51.6 & -44:32:17.5 & 1.8 & 4.1$\pm$0.6 & 0.6$\pm$0.4 & -2.37 & 0.6	$\pm$	0.2	&	-0.2	$\pm$	0.2	&	-0.3	$\pm$	0.2	&	-0.4	$\pm$	0.4	&fgstar\\
43 & 01:51:52.1 & -44:25:46.1 & 1.6 & 10.5$\pm$0.6 & 1.3$\pm$0.2 &  & 0.13	$\pm$	0.07	&	0.00	$\pm$	0.07	&	-0.32	$\pm$	0.09	&	-1.0	$\pm$	0.2	&[AGN]\\
80 & 01:51:55.4 & -44:32:22.5 & 1.6 & 15$\pm$1 & 4.0$\pm$0.7 & 0.24 & 0.79	$\pm$	0.08	&	0.21	$\pm$	0.08	&	-0.13	$\pm$	0.08	&	0.0	$\pm$	0.1	&[hard]\\
60 & 01:51:56.0 & -44:29:28.7 & 1.6 & 10$\pm$1 & 3.1$\pm$0.7 & -0.56 & 0.7	$\pm$	0.1	&	-0.2	$\pm$	0.1	&	-0.1	$\pm$	0.2	&	0.4	$\pm$	0.1	&[gal]\\
83 & 01:51:57.2 & -44:32:55.3 & 1.6 & 16$\pm$1 & 3.9$\pm$0.8 &  & 0.70	$\pm$	0.09	&	0.30	$\pm$	0.06	&	-0.20	$\pm$	0.07	&	-0.3	$\pm$	0.2	&[hard]\\
30 & 01:51:57.4 & -44:23:06.1 & 1.6 & 18$\pm$1 & 5.7$\pm$0.7 & -0.20 & 0.09	$\pm$	0.08	&	0.11	$\pm$	0.07	&	-0.25	$\pm$	0.08	&	0.2	$\pm$	0.1	&[hard]\\
57 & 01:52:00.7 & -44:29:14.6 & 1.6 & 8.5$\pm$0.8 & 4.0$\pm$0.7 &  & 1.0	$\pm$	0.2	&	0.5	$\pm$	0.1	&	0.23	$\pm$	0.09	&	-0.1	$\pm$	0.1	&[hard]\\
97 & 01:52:00.9 & -44:35:54.5 & 1.6 & 25$\pm$1 & 5.0$\pm$0.9 &  & 0.37	$\pm$	0.07	&	-0.10	$\pm$	0.06	&	-0.08	$\pm$	0.07	&	-0.2	$\pm$	0.1	&[hard]\\
19 & 01:52:02.6 & -44:22:08.7 & 1.6 & 12.1$\pm$0.9 & 3.0$\pm$0.6 &  & 0.5	$\pm$	0.1	&	0.30	$\pm$	0.08	&	-0.08	$\pm$	0.08	&	-0.2	$\pm$	0.1	&[hard]\\
26 & 01:52:11.1 & -44:22:56.2 & 1.7 & 5.0$\pm$0.6 & 5$\pm$1 &  & 1.0	$\pm$	1.9	&	0.96	$\pm$	0.07	&	0.2	$\pm$	0.1	&	-0.5	$\pm$	0.2	&[hard]\\
5 & 01:52:11.3 & -44:18:08.8 & 1.6 & 27$\pm$3 & 31$\pm$6 &  & 0.6	$\pm$	0.1	&	0.4	$\pm$	0.1	&	-0.3	$\pm$	0.1	&	0.3	$\pm$	0.2	&[hard]\\
37 & 01:52:25.0 & -44:25:16.5 & 1.7 & 9.1$\pm$0.9 & 11$\pm$2 &  & 0.2	$\pm$	0.2	&	0.5	$\pm$	0.1	&	-0.2	$\pm$	0.1	&	0.3	$\pm$	0.1	&[hard]\\
\hline
\end{tabular}
\end{table}
\end{landscape}
\end{document}